\documentclass[11pt]{article}
\usepackage[utf8]{inputenc}
\usepackage{enumerate}
\usepackage{xcolor}
\usepackage{amsmath}
\usepackage{amsfonts}
\usepackage{amssymb}
\usepackage{capt-of}
\usepackage[a4paper]{geometry}
\usepackage{float}
\usepackage{graphicx}
\usepackage{authblk}
\usepackage{authblk}
\usepackage{graphics, setspace}
\usepackage[title]{appendix}
\usepackage{hyperref}
\newcommand{\mathsym}[1]{{}}
\newcommand{\unicode}[1]{{}}

\begin{document}

\providecommand{\abs}[1]{\lvert#1\rvert}
\newcommand\ii{\'{\i}}
\newcommand\oo{\'{\o}}
\title{Casimir energy in 2+1 dimensional field theories\\}
\author[1]{Manuel Asorey}
\author[2]{Claudio Iuliano}
\author[1]{Fernando Ezquerro}

\affil[1]{\ Departamento de F\'isica Te\'orica, Centro de Astropart\'{\i}culas y F\'{\i}sica de Altas, Energ\'{\i}as, Universidad de Zaragoza, 50009 Zaragoza, Spain}
\affil[2]{\ Max Planck Institute for Mathematics in Sciences (MiS), Inselstraße 22, 04103 Leipzig, Germany}

\date{}
\maketitle

\begin{abstract}
	\normalsize
We explore the dependence of vacuum energy on the boundary conditions for massive scalar fields in (2 + 1)-dimensional spacetimes. We consider the simplest geometrical setup given by a two-dimensional space bounded by two homogeneous parallel wires in order to compare it with the non-perturbative behaviour of the Casimir energy for non-Abelian gauge theories in (2 + 1) dimensions. Our results show the existence of two types of boundary conditions which give rise to two different asymptotic exponential decay regimes of the Casimir energy at large distances. The two families are distinguished by the feature that the boundary conditions involve or not interrelations between the behaviour of the fields at the two boundaries. Non-perturbative numerical simulations and analytical arguments show such an exponential decay for Dirichlet boundary conditions of SU(2) gauge theories. The verification that this behaviour is modified for other types of boundary conditions requires further numerical work. Subdominant corrections in the low-temperature regime are very relevant for numerical simulations, and they are also analysed in this paper.\\
\end{abstract}
\providecommand{\keywords}[1]
{
	\small	
	\textbf{Keywords:} #1
}

\newpage
\section{Introduction}

The role of boundary effects in quantum field theory is fundamental for many quantum phenomena. One of the earliest applications was the Casimir effect  \cite{Casimir:1948dh}. A quantum field, when confined between two solid bodies, generates a dependence of the renormalized vacuum energy on the boundary conditions at the interfaces of the bodies. This dependence of the vacuum energy generates a force between them which  depends on the nature of the boundary conditions of the quantum fields. Although this effect is very tiny, it has been  experimentally measured  in different setups. \cite{sparnaay1957attractive,sparnaay1958measurements,lamoreaux1997demonstration,PhysRevLett.81.4549,PhysRevLett.87.211801, Dalvit}. 

A remarkable effort has been made in understanding and computing the Casimir effect for different models and setups.  Some relevant results were obtained in Ref. \cite{Boundary_general_2013}, where the Casimir vacuum energy at zero temperature was computed for general boundary conditions and arbitrary dimensions for massless scalar fields using heat kernel methods. These results were later extended to finite temperatures in 3+1 dimensions \cite{munoz2020thermal}.

Less known are the characteristics of the effect for interacting theories \cite{SYMANZIK19811}. Quite recently, the behaviour of the Casimir energy has been investigated in 2+1 dimensional Yang-Mills theories, where  some reparametrization of gauge fields in terms of scalar fields allows an analytic approach to the problem \cite{KARABALI1998661,PhysRevD.98.105009,karabali1996origin}. Numerical simulations with Dirichlet boundary conditions on  gauge fields confirm the results of this analytic approach \cite{PhysRevLett.121.191601}. 

For $SU(2)$ gauge theories the analytic approach is based on the description of gauge fields
in terms of a massive  scalar field, whose mass depends on the gauge coupling that in 2+1 dimensions 
is not dimensionless as in 3+1 dimensions. In that case, the Casimir energy of the strongly interacting gauge theories with Dirichlet boundary conditions does coincide with the Casimir energy of a scalar field with a magnetic mass $m=\frac{g^2}{\pi}$, where $g$ is the gauge coupling constant.

Some numerical simulations are in progress  with different boundary conditions on gauge fields \cite{asorey2023energy} to test if the relation between Casimir energy of massive fields and Yang Mills theory is robust under the change of boundary conditions. Making comparisons with what happens for scalar fields requires to know  the behaviour of Casimir energy of the massive scalars for different families of boundary conditions.

In this paper, we study the vacuum energy for massive scalar fields with general boundary conditions in a two dimensional setup bounded by  two homogeneous parallel wires by using a regularization scheme similar to the one used in \cite{asorey_temp1,asorey2015topological} for massless theories.  To compare with the lattice gauge theories results, it is necessary to work at finite temperature; thus, it is important to understand how the thermal fluctuations affect the Casimir energy at low temperatures both in 2+1 dimensional $SU(2)$ gauge theories and massive scalar field theories in order to have some analytical background reference to results to.

Independently of these motivations, some interest has been raised recently on the applications of the thermal Casimir effect in nano-electronic devices \cite{zou2013casimir, terccas2017quantum} or the appearance of negative self entropy related to this effect \cite{geyer2005thermal,milton2017casimir,bordag2018free} which boosted the interest on new aspects of these thermal effects.

\section{Effective action of a Massive Scalar Field in 2+1 Dimensions}

We consider a free scalar massive field in 2+1 dimensions confined between two homogeneous infinite wires separated by a distance $L$. Depending on the structure of the wires the quantum fields have to satisfy some conditions on the boundary wires. Moreover, finite temperature $T\neq 0$ effects can be described in  the Euclidean formalism by compactification of the Euclidean time direction into a  circle of radius $\frac{\beta}{2\pi}=\frac1{2\pi T}$. In this case, the partition function can be written as the following determinant
\begin{equation}\label{det}
	Z(\beta)=\text{det}\left(-\partial_0^2-\boldsymbol{\nabla}^2+m^2\right)^{-1/2},
\end{equation}
where $m$ is the mass of the fields, $\boldsymbol{\nabla}^2$ the spatial Laplacian and $\partial_0$ the Euclidean time derivative. As already mentioned, the boundary conditions are periodic in time 
$\psi(t+\beta,\mathbf{x})=\psi(t,\mathbf{x})$, 
and because of the homogeneity of the boundary wires  the spatial boundary 
conditions can be given in terms of of  $2\times 2$ unitary matrices 
\textcolor{black}{$U\in$ U(2)} \cite{asorey2005global} 
\begin{equation}
	\varphi - i\delta\dot\varphi = U(\varphi +i\delta\dot\varphi ),
	\label{bccc}
\end{equation}
where
$\delta$ is an arbitrary scale parameter and
\begin{equation}
	\varphi=\begin{pmatrix}
		\varphi(L/2)  \\ \varphi(-L/2)  \\ 
	\end{pmatrix},
	\quad
	\dot\varphi=\begin{pmatrix}
		\dot\varphi(L/2)  \\ \dot\varphi(-L/2)  \\ 
	\end{pmatrix},
\end{equation}
are the boundary values  $\varphi(\pm L/2)= \psi(t,x_1, \pm  L/2)$ of  
the fields  $\psi$   and  their outward normal derivatives 
$\dot\varphi(\pm L/2)=\pm\partial_2\psi(t,x_1, \pm L/2)$   
on the wires. From now on we will assume that $\delta=1$ for simplicity.

In the standard parametrization of U(2) matrices
\begin{eqnarray}\label{U2}
	U(\alpha,\gamma,{{\bf n}})&=&{\mathrm e}^{i\alpha}\left(\mathbb{I }\cos\gamma+i{{\bf n}}\cdot\boldsymbol{\sigma}\,\sin\gamma \right);
	\label{parametrization}\quad  {\alpha\in[0,2\pi],\,\, \gamma\in[-\pi/2,\pi/2]}
\end{eqnarray}
in terms of an unit vector ${{\bf n}}\in S^2$ and Pauli matrices $\boldsymbol{\sigma}$,
the space of boundary conditions that preserve
the non-negativity of the spectrum of the operator $-\boldsymbol{\nabla}^2$
is restricted by the inequalities  \textcolor{black}{$0\leq\alpha\pm\gamma\leq\pi$ }\footnote{Moreover, since the scalar field is real, the second component of the unit vector n has to vanish, i.e., $n_2=0$}
.\cite{Boundary_general_2013}.

The determinant of the second order differential operator $-\partial_0^2-\boldsymbol{\nabla}^2+m^2$ in equation (\ref{det}) is ultraviolet (UV) divergent but can be regularized by means of zeta regularization method \cite{PhysRevD.13.3224,Blau:1988kv}. The effective action is defined by the logarithm of the partition function which can be expressed as 
\begin{equation}
	S_{\text{eff}}=-\log Z=-\frac{1}{2}\frac{d}{ds}\zeta \left(s\right)|_{s=0},
\end{equation}
in terms of the zeta function 
\begin{equation}
	\zeta(L,m,\beta;s)= \mu^{2s}\left(-\partial_0^2-\boldsymbol{\nabla}^2+m^2\right)^{-s}
\end{equation}
where we have introduced the scale parameter $\mu$, which encodes the standard ambiguity of zeta function renormalization techniques \footnote{See e.g. \cite{PhysRevD.56.7797,10.1063/1.532929} for a detailed discussion and comparison with other renormalization methods.}, to make the  zeta function  dimensionless. This ambiguity will be fixed by the renormalization scheme prescription. Actually, the scale parameter $\mu$ can be seen as an explicit implementation of the renormalization group. 

In our case of a massive scalar field confined between two infinite wires, the eigenvalues of operator $-\partial_0^2-\boldsymbol{\nabla}^2$ are given by the sum of the square of the temporal modes $2\pi l/\beta$ associated to the Matsubara frequencies, the continuous spatial modes $k$ and the discrete spatial modes $k_i$ that depend on the boundary conditions imposed by the boundary wires
\begin{equation}
	\lambda= \left(\frac{2\pi l}{\beta}\right)^2+k^2+k_i^2+m^2 \hspace{2cm}l \in \mathbb Z, i\in {\mathbb N}.
\end{equation} 
Thus, the zeta function  in this case reads as follows
\begin{equation}\label{ini_1}
	\zeta(L,m,\beta;s)=\mu^{2s}\frac{A}{2\pi}\sum_{l,i} \int_{-\infty}^\infty dk\left(\left(\frac{2\pi l}{\beta}\right)^2+k^2+k_i^2+m^2\right)^{-s},
\end{equation}
where $A$ is the length of the wires. Now we can integrate the continuous spatial modes using the analytic extension of the zeta function
\begin{equation}\label{zeta_2d}
	\zeta(L,m,\beta;s)=\mu^{2s}\frac{A\Gamma(s-1/2)}{2\sqrt \pi\Gamma(s)}\sum_{l,i} \left(\left(\frac{2\pi l}{\beta}\right)^2+k_i^2+m^2\right)^{-s+1/2} .
\end{equation}
It was shown in \cite{Boundary_general_2013} that for homogeneous boundary conditions along the wires  the discrete spatial modes are given by the zeros of   the spectral function
\begin{equation}\label{spectral_2d}
	h^L_U(k)=2i\left(\sin(kL)\left((k^2-1)\cos \gamma +(k^2+1)\cos \alpha \right)-2k\sin \alpha \cos (kL)-2kn_1\sin \gamma\right),
\end{equation}
in the following way
\begin{equation}\label{zeta_integral}
	\zeta(L,m,\beta;s)=\mu^{2s}\frac{A\Gamma(s-1/2)}{4\pi^{3/2}i\Gamma(s)}\sum_{l=-\infty}^\infty \oint dk \left(\left(\frac{2\pi l}{\beta}\right)^2+k^2+m^2\right)^{-s+1/2}\frac{d}{dk}\log h^L_U(k),
\end{equation}
where  the integral \textcolor{black}{ is defined along the contour of }a thin strip enclosing the positive real axis where all the zeros of the spectral function $h_U(k)$ are located. 

All ultraviolet divergences arise in the zero temperature limit 
of the vacuum energy and the removal of such a divergences require a consistent prescription  method ({\sl renormalization scheme}) with a clear physical meaning. They appear in the leading terms of the zero temperature expansion that has the following asymptotic behaviour in the large $L$ limit \cite{Boundary_general_2013,asorey_temp1}
\begin{equation}
		S_{\text{eff}}^{l=0}= \beta E_0 =
		C_0(m)A\ \beta L+C_1(m)A\beta+\frac{A\beta}{L}C_c(m,L)+\ldots. 
\end{equation}
where $E_0$ is the vacuum energy, $C_0(m)$ the divergent bulk vacuum energy density, $C_1(m)$  the divergent energy density of the boundary wires and $C_c(m,L)$ is the the finite coefficient of the Casimir energy.

One renormalization prescription which allows us to get rid of all these divergences consists on the redefinition of  the renormalized effective action as \cite{asorey_temp1, asorey2015topological} 
\begin{equation}\label{zeta_combination}
	S^{\text{ren}}_{\text{eff}}=-
	\frac{1}{2}\frac{d}{ds} \zeta_{\text{ren}}
	(L,m,\beta;s) 
	\Bigr|_{s=0}\ ,
\end{equation}
where
\begin{equation}\label{zeta_re}
	\zeta_{\text{ren}}(L,m,\beta;s)=\lim_{L_0\rightarrow \infty}\left(\zeta(L,m,\beta;s)+\zeta(2L_0+L,m,\beta;s)-2\zeta(L_0+L,m,\beta;s)\right),
\end{equation}
in terms of an auxiliary length $L_0$. Notice that the physical condition which fixes this renormalization scheme is the complete removal of the spurious contributions to the bulk  and the boundary vacuum energies, leaving only Casimir energy  terms and  non linear in $\beta$  temperature dependent contributions to the effective action. These are precisely  the physical requirements that fix the renormalization scheme prescription.

The sum over Matsubara modes can be explicitly computed in the low temperature regime. 

\section{Low Temperature regime}
In the low temperature limit  $\beta m\gg  1$, we can not  express the result as an infinite series of $\frac1{\beta}$. This means that we have to deal first with the Matsubara modes and later with the boundary modes. We start by rewriting \eqref{zeta_2d} as follows
\begin{equation}
	\zeta(L,m,\beta;s)=\left(\frac{\mu \beta}{2\pi}\right)^{2s}\frac{A\sqrt \pi \Gamma(s-1/2)}{\beta \Gamma(s)}\sum_{i}\sum_{l=-\infty}^\infty \left( l^2+\left(\frac{k_i \beta}{2\pi}\right)^2+\left(\frac{m \beta}{2\pi}\right)^2\right)^{-s+1/2}.
\end{equation}

Now, we use the Mellin transform
\begin{equation}\label{Mellin_t}
	\zeta(L,m,\beta;s)=\left(\frac{\mu \beta}{2\pi}\right)^{2s}\frac{A\sqrt \pi}{\beta \Gamma(s)}\sum_{i}\sum_{l=-\infty}^\infty\int_0^\infty dt\ t^{s-3/2}\	e^{-\left( l^2+\left(\frac{k_i \beta}{2\pi}\right)^2+\left(\frac{m \beta}{2\pi}\right)^2\right)t}
\end{equation}
and apply the Poisson formula for the sum over $l$ modes
\begin{equation}\label{Poisson}
	\zeta(L,m,\beta;s)=\left(\frac{\mu \beta}{2\pi}\right)^{2s}\frac{A \pi}{\beta \Gamma(s)}\sum_{i}\sum_{l=-\infty}^\infty\int_0^\infty dt\ t^{s-2}\	e^{-\left(\left(\frac{k_i \beta}{2\pi}\right)^2+\left(\frac{m \beta}{2\pi}\right)^2\right)t-\frac{(\pi l)^2}{t}}.
\end{equation}

We can compute the integral
\begin{align}\nonumber
	\zeta(L,m,\beta;s)&=\left(\frac{\mu \beta}{2\pi}\right)^{2s}\frac{A\pi}{\beta \Gamma(s)}\left(\Gamma \left(s-1\right)\sum_i\left(\left(\frac{k_i \beta}{2\pi}\right)^2+\left(\frac{m \beta}{2\pi}\right)^2\right)^{1-s}+\right.\\
	&\left. +4\sum_{i}\sum_{l=1}^\infty\left(\pi l\right)^{-1+s}\left(\left(\frac{k_i \beta}{2\pi}\right)^2+\left(\frac{m \beta}{2\pi}\right)^2\right)^{1/2-s/2}K_{1-s}\left(\beta l\sqrt{k_i^2+m^2}\right)\right),
\end{align}
where we have obtained a term ($l=0$) that has a linear dependence on $\beta$,   and the rest of  terms can be  expressed in terms of  the modified Bessel function of second type $K_\nu$. 
Let us focus on the first term which is the zero temperature one, by replacing the sum of boundary modes by an integral modulated by the spectral function \eqref{spectral_2d} we have
\begin{equation}
	\zeta^{l=0}(L,m,\beta;s)=\mu^{2s}\frac{A\beta  \Gamma \left(s-1\right)}{8\pi^2 i \Gamma(s)}\oint dk\ \left(k^2+m^2\right)^{1-s}\frac{d}{dk}\log h^L_U(k).
\end{equation}
Thus, the zero temperature term  of the renormalized zeta \eqref{zeta_re} is
\begin{equation}
	\zeta_{\text{ren}}^{l=0}(L,m,\beta;s)=\mu^{2s}\frac{A\beta  \Gamma \left(s-1\right)}{8\pi^2 i \Gamma(s)}\lim_{L_0\rightarrow \infty}\oint dk \left(k^2+m^2\right)^{1-s}\frac{d}{dk}\log \frac{h^L_U(k)h^{2L_0+L}_U(k)}{\left(h^{L_0+L}_U(k)\right)^2}.
\end{equation}
As it was explained previously, this combination cancels the UV divergences on the integral, thus, the only divergent terms left are the ratio of two Gamma functions whose asymptotic behaviour in  the small $s$ expansion is
\begin{equation}\label{s_expansion_G}
	\frac{\Gamma(s-1)}{\Gamma(s)}=-1-s+\mathcal O(s^2)
\end{equation}  which allows to calculate the derivative 
\begin{equation}
	\frac{d}{ds}\left.\left(-\left(1+s\right)\left(k^2+m^2\right)^{1-s}\mu^{2s}\right)\right|_{s=0}=(k^2+m^2)\left(\log\left(k^2+m^2\right)-2\log \mu-1\right).
\end{equation}
Thus, we have
\begin{align}\nonumber
	\left(\zeta_{\text{ren}}^{l=0}\right)'(L,m,\beta;0)=\frac{A \beta }{8\pi^2 i }\lim_{L_0\rightarrow \infty}\oint dk &\left(k^2+m^2\right)\left(\log\left(k^2+m^2\right)-2\log \mu-1\right)\\
	&\times\left(\frac{d}{dk}\log \frac{h^L_U(k)h^{2L_0+L}_U(k)}{\left(h^{L_0+L}_U(k)\right)^2}\right).
\end{align}
Since the integrand is holomorphic, we can extend the integration contour to an infinite semicircle limited by the imaginary axis on its left. Also, because the integration over the semicircle is zero, we can reduce the integral to the imaginary axis
\begin{align}\nonumber
	\left(\zeta_{\text{ren}}^{l=0}\right)'(L,m,\beta;0)=\frac{A \beta }{8\pi^2i }\lim_{L_0\rightarrow \infty}\int_{-\infty}^\infty  dk\left(k^2-m^2\right)
	&\left(\log\left(m^2-k^2\right)-2\log \mu-1\right)\\
	&\times\left(\frac{d}{dk}\log \frac{h^L_U(ik)h^{2L_0+L}_U(ik)}{\left(h^{L_0+L}_U(ik)\right)^2}\right).
\end{align}

Taking into account that the integrand is parity odd, the integral would vanish if it were not by the contribution of the branching point $k=m$ of the logarithm $\log(m^2-k^2)$, which gives a factor $2\pi i$ for the interval $(m,\infty)$ which is absent in the interval $(-\infty,-m)$. Thus, the expression reduces to
\begin{equation}
	\left(\zeta_{\text{ren}}^{l=0}\right)'(L,m,\beta;0)=\frac{A\beta }{4\pi }\lim_{L_0\rightarrow\infty}\int_{m}^\infty  dk \left(k^2-m^2\right)\frac{d}{dk}\log \frac{h^L_U(ik)h^{2L_0+L}_U(ik)}{\left(h^{L_0+L}_U(ik)\right)^2}.
\end{equation}
Since the integral domain begins at $m$, we can take the limit $L_0\rightarrow\infty$ on the spectral functions by noticing that 
\begin{equation}\label{L0 infinity}
	\lim_{L_\ast\rightarrow \infty} h_U^{L_\ast}(ik)=\lim_{L_\ast\rightarrow \infty}e^{k(L_\ast)}\left((k^2+1)\cos \gamma +(k^2-1)\cos \alpha +2k\sin\alpha \right).
\end{equation}

If we define the result in terms of the limit for the spectral function 
\begin{equation}
	h^\infty_U(ik)\equiv \left((k^2+1)\cos \gamma +(k^2-1)\cos \alpha +2k\sin\alpha \right),
\end{equation} we get a simplified formula to
\begin{equation}
	\left(\zeta_{\text{ren}}^{l=0}\right)'(L,m,\beta;0)=-\frac{A\beta }{4\pi }\int_{m}^\infty  dk \left(k^2-m^2\right)\left(L-\frac{d}{dk}\log \frac{h^L_U(ik)}{h^{\infty}_U(ik)}\right).
\end{equation}

\subsection{Temperature dependent terms}
Let us now compute the terms with $l\not =0$ of the zeta function
\begin{align} \nonumber
	\zeta^{l\not =0}(L,m,\beta;s)=\left(\frac{\mu \beta}{2\pi}\right)^{2s}\frac{4A\pi}{\beta \Gamma(s)} \sum_{i,l=1}\left(\pi l\right)^{-1+s}&\left(\left(\frac{k_i \beta}{2\pi}\right)^2+\left(\frac{m \beta}{2\pi}\right)^2\right)^{1/2-s/2}\\
	&\times\left(K_{1-s}\left(\beta l\sqrt{k_i^2+m^2}\right)\right).
\end{align}
Since the Bessel special function $K_1$ decreases exponentially as the argument grows, both sums are finite, thus, the only divergent contribution is the Gamma function, which after derivation gives
\begin{equation}
	\left(\zeta^{l\not =0}\right)'(L,m,\beta;0)=\frac{2A}{\pi }\sum_i\sum_{l=1}^\infty\frac{\sqrt{k_i^2+m^2}}{l}K_{1}\left(\beta l\sqrt{k_i^2+m^2}\right)
\end{equation}
and we can we rewrite the sum of the discrete eigenvalues by means of the spectral formula \eqref{spectral_2d}
\begin{equation} 
	\left(\zeta^{l\not =0}\right)'(L,m,\beta;0)=\frac{A}{\pi^2 i}\sum_{l=1}^\infty\oint dk\ \frac{\sqrt{k^2+m^2}}{l}K_{1}\left(\beta l\sqrt{k^2+m^2}\right) \frac{d}{dk}\log\left(h^L_U(k)\right).
\end{equation}
Thus, the temperature dependent terms of the renormalized zeta function \eqref{zeta_re} have the following form
\begin{equation} 
	\left(\zeta_{\text{ren}}^{l\not =0}\right)'(L,m,\beta;0)=\frac{A}{\pi^2 i}\lim_{L_0\rightarrow\infty}\sum_{l=1}^\infty\oint dk\ \frac{\sqrt{k^2+m^2}}{l}K_{1}\left(\beta l\sqrt{k^2+m^2}\right) \frac{d}{dk}\log \frac{h^L_U(k)h^{2L_0+L}_U(k)}{\left(h^{L_0+L}_U(k)\right)^2}.
\end{equation}
In a similar manner as we did for the $l=0$ term, since the integrand is also holomorphic  we can extend the contour to an infinite semi-circle limited by the imaginary axis. Because the integral is zero on the semicircle, we can reduce the integral  just to the imaginary axis 
\begin{align} \nonumber
	\left(\zeta_{\text{ren}}^{l\not =0}\right)'(L,m,\beta;0)=-\frac{A}{\pi^2 i}\lim_{L_0\rightarrow\infty}\sum_{l=1}^\infty\int_{-\infty}^\infty &dk\ \frac{\sqrt{-k^2+m^2}}{l}K_{1}\left(\beta l\sqrt{-k^2+m^2}\right)\\
	&\times\left(\frac{d}{dk}\log \frac{h^L_U(ik)h^{2L_0+L}_U(ik)}{\left(h^{L_0+L}_U(ik)\right)^2}\right).
\end{align}
Because the integrand is odd, the contribution of $(-m,m)$ is zero, whereas the branching point of $\sqrt{m^2-k^2}$ introduces a change of sign on the integrand on $(-\infty,-m)$ and also in the argument of the Bessel function. Using that $K_1(\bar z)=\overline{ K_1(z)}$, the real part of the integrals between $(-\infty,-m)$ and $(m,\infty)$ is twice one of the integrals, whereas the imaginary part cancels out. In summary,  the integral  can be reduced to
\begin{align} \nonumber
	\left(\zeta_{\text{ren}}^{l\not =0}\right)'(L,m,\beta;0)=-\frac{2A}{\pi^2}\lim_{L_0\rightarrow\infty}\sum_{l=1}^\infty\int_{m}^\infty &dk\ \frac{\sqrt{k^2-m^2}}{l}\Re \left(K_{1}\left(i\beta l\sqrt{k^2-m^2}\right)\right)\\
	&\times\left(\frac{d}{dk}\log \frac{h^L_U(ik)h^{2L_0+L}_U(ik)}{\left(h^{L_0+L}_U(ik)\right)^2}\right).
\end{align}

We can take the limit $L_0\rightarrow\infty$  using \eqref{L0 infinity} as
we did for the $l=0$ term and the integral is simplified to
\begin{equation} 
	\left(\zeta_{\text{ren}}^{l\not =0}\right)'(L,m,\beta;0)=\frac{2A}{\pi^2}\sum_{l=1}^\infty\int_{m}^\infty dk\ \frac{\sqrt{k^2-m^2}}{l}\Re \left(K_{1}\left(i\beta l\sqrt{k^2-m^2}\right)\right)\left(L-\frac{d}{dk}\log \frac{h^L_U(ik)}{h_U^\infty(ik)}\right).
\end{equation} 

\section{Casimir energy}
The Casimir energy can be derived from the terms we have just computed in the previous sections. 
We can easily compute the free energy with the effective action simply as  $F=S_{\text{eff}}/\beta$. This free energy has two different contributions \cite{bordag2018free}, the non temperature dependent part ($l=0$) which corresponds to the Casimir energy of the system  
\begin{equation}\label{Cas_spectral}
	F^{l=0}_U(L,m,\beta)=E_U^c(L,m)=\frac{A}{8\pi }\int_{m}^\infty  dk \left(k^2-m^2\right)\left(L-\frac{d}{dk}\log \frac{h^L_U(ik)}{h^{\infty}_U(ik)}\right),
\end{equation} and the temperature dependent part 
\begin{equation}\label{Free_spectral}
	F^{l\not =0}_U(L,m,\beta)=-\frac{A}{\beta \pi^2}\sum_{l=1}^\infty\int_{m}^\infty dk\ \frac{\sqrt{k^2-m^2}}{l}\Re \left(K_{1}\left(i\beta l\sqrt{k^2-m^2}\right)\right)\left(L-\frac{d}{dk}\log \frac{h^L_U(ik)}{h_U^\infty(ik)}\right).
\end{equation}
Both terms of the free energy decrease to zero as the distance between the wires $L$ grows to infinite, which is the  expected behaviour. The temperature dependent term also vanishes $F^{l\not =0}_U\rightarrow 0$ when the temperature also does ($\beta \rightarrow\infty$).

\subsection{Asymptotic behaviour }
Let us now analyse the behaviour of the Casimir energy when $mL\rightarrow \infty$. First, we rewrite the hyperbolic functions of the spectral function as
\begin{equation}\nonumber
	h_U^L(ik)=e^{kL}\left((k^2+1)\cos\gamma+(k^2-1)\sin\alpha+2k\sin\alpha\right)\left(1+n_1\sin(\gamma) {\mathcal {A}}\ e^{-kL}+{\mathcal {B}}\ e^{-2kL}\right),
\end{equation}
where ${\mathcal {A}}$ and ${\mathcal {B}}$ are
\begin{align}
	&{\mathcal {A}}(k,\alpha,\gamma)=\frac{4k}{(k^2+1)\cos\gamma+(k^2-1)\sin\alpha+2k\sin\alpha}\\
	&{\mathcal {B}}(k,\alpha,\gamma)=\frac{-(k^2+1)\cos\gamma-(k^2-1)\sin\alpha+2k\sin\alpha}{(k^2+1)\cos\gamma+(k^2-1)\sin\alpha+2k\sin\alpha}.
\end{align}

We can use this expression to approximate the logarithm of the quotient of spectral functions as
\begin{equation} 
	\log \frac{h_U^L(ik)}{h_U^\infty(ik)}=kL+n_1\sin\gamma {\mathcal {A}}\ e^{-kL}+({\mathcal {B}}-\frac{{\mathcal {A}}'}{2})e^{-2kL}+O(e^{-3kL}),
\end{equation}
where ${\mathcal {A}}'=(n_1\sin(\gamma){\mathcal {A}})^2$, and we expanded the logarithm in powers of $e^{-kL}$. Now we  introduce this expression on the integral of the Casimir energy formula
\begin{align}\nonumber
	E_U^c&=-\frac{A}{8\pi}\int_m^\infty dk (k^2-m^2)\frac{d}{dk}\left(n_1\sin\gamma {\mathcal {A}}\ e^{-kL}+({\mathcal {B}}-\frac{{\mathcal {A}}'}{2})e^{-2kL}+O(e^{-3kL})\right)\\
	&=\frac{A}{4\pi}\int_m^\infty dk\ k\left(n_1\sin\gamma {\mathcal {A}}\ e^{-kL}+({\mathcal {B}}-\frac{{\mathcal {A}}'}{2})e^{-2kL}+O(e^{-3kL})\right).
\end{align}

We can expand this integral as a power series of $1/L$ for each exponential order by integrating by parts as follows
\begin{equation} 
	\int_m^\infty dk g(\alpha,\gamma,n_1,k) e^{-jkL}=-\left.\frac{g(\alpha,\gamma,n_1,k)}{jL}e^{-jkL}\right|_m^\infty+\int_m^\infty dk\  \frac{g(\alpha,\gamma,n_1,k)'}{jL} e^{-jkL},
\end{equation}
and iterate this process since all  derivatives of $g(\alpha,\gamma,n_1,k)$ are regular in $[m,\infty]$. Thus, the Casimir energy is given by
\begin{equation}
	E_U^c=\sum_{j=1}^\infty\sum_{\nu=1}^\infty \frac{c_{j,\nu}(\alpha,\gamma,n_1,m)}{(jL)^\nu}e^{-jmL},
\end{equation}
where the coefficients corresponding to the leading order in the exponential expansion are of the form
\begin{equation}
	c_{1,\nu}=-\frac{n_1\sin\gamma}{4\pi}\left.\frac{d^\nu(k{\mathcal {A}}(\alpha,\gamma,k))}{dk^{\nu}}\right|_m^\infty.
\end{equation}
This means that when $n_1\sin\gamma=0$  all the terms that behave as $e^{-mL}$ vanish and the leading contribution will be of the order of $e^{-2mL}$. Thus, we have two different families of \textcolor{black}{boundary} conditions with different asymptotic behaviour
\begin{equation}\label{rate}
	L E_U^c\sim \left\{ \begin{matrix}e^{-mL} &\text{if  }\text{ tr}(U\sigma_1)\not =0\\
		e^{-2mL} &	\text{if  }\text{ tr}(U\sigma_1)=0,  \end{matrix} \right.
\end{equation}
depending on whether the matrix $U$ that defines the boundary conditions depends or not on $\sigma_1$.

This is the most important result of this paper because
it classifies the boundary conditions in two families. The difference between the two families is  the rate of the exponential decay of the Casimir energy \eqref{rate}. 

The physical characterization of the two families of boundary conditions with different exponential decays 
is the vanishing or not of $\text{ tr}(U\sigma_1)$. The non-vanishing case corresponds to boundary conditions 
which connect the values of the fields or its normal derivatives  at the two boundary wires, whereas the vanishing
case corresponds to families of boundary conditions which only constraints the values of the fields or its normal derivatives 
at each boundary separately.

The result is obtained for a massive free field bosonic theory.
If the observed rate of decay in gauge theories has the same behaviour it will provide a strong evidence of the scenario that
describes the dynamics of gauge theories in 2+1 dimensions in terms of a \textcolor{black}{bosonic} massive scalar field.

$\,$
\bigskip
$\,$

\bigskip

\section{ Special cases of boundary conditions}
Let us analyse  some particular cases where the integral of the Casimir energy can be analytically computed and which are of special interest for their potential implementation for gauge fields. {An alternative derivation based on the explicit calculation of the spectrum of spatial Laplacian for these cases is postponed to Appendix \ref{app1}}.\\

\subsection{Dirichlet and Neumann boundary conditions}

Dirichlet boundary conditions corresponds to the physical case of fields vanishing at both boundary wires $\varphi(L/2)=\varphi(-L/2)=0$; in our parametrization \eqref{U2} they are given by $U_D=-\mathbb{I}$. 
Notice that these  boundary conditions do not relate the boundary values of the fields of one 
boundary with the boundary values at the other one. 

The derivative of the logarithm of the quotient of spectral functions is
\begin{equation}\label{spectra_dir}
	\frac{d}{dk}\log\left(h^L_{U_D}(ik)/h^\infty_{U_D}(ik)\right)=L\coth (kL).
\end{equation}

We can integrate the Casimir energy formula \eqref{Cas_spectral}
\begin{align}\label{Cas_Dir}
	E^c_{D}(L,m)=-\frac{A}{16\pi L^2}\left(2mL\ \text{Li}_2\left(e^{-2mL}\right)+\text{Li}_3\left(e^{-2mL}\right)\right),
\end{align}
which, in the massless limit gives
\begin{equation}
	E^c_D(L,0)=-\frac{A\zeta(3)}{16\pi L^2}.
\end{equation}

But in the asymptotic very large  $mL\gg 1$ limit the Casimir energy  has a fast
exponential decay $e^{-2 mL}$ as predicted by the fact that 
$-\mathrm{Tr}\, \mathbb{I}\, \sigma_1= 0$.

The temperature dependent terms of the free energy
\begin{equation}\label{Free_Dir}
	F^{l\not =0}_{D}(L,m,\beta)=-\frac{AL}{\beta\pi^2}\sum_{l=1}^\infty\int_{m}^\infty dk\ \frac{\sqrt{k^2-m^2}}{l}\Re \left(K_{1}\left(i\beta l\sqrt{k^2-m^2}\right)\right)\left(1-\coth (kL)\right)
\end{equation}
cannot be analytically  computed but from the asymptotic expansion of the term $$1-\coth (kL)\approx -e^{-2kL}$$ of the integrand, it can be shown that they have the same exponential decay \textcolor{black}{ with $mL$ as the Casimir energy \eqref{Cas_Dir}}.

Neumann boundary conditions  correspond to the case where normal derivative 
of the fields vanish at the boundary wires $\dot\varphi(L/2)=\dot\varphi(-L/2)=0$. 
They are parameterized by the 
unitary matrix $U_N=\mathbb I$. The derivative of the logarithm of the quotient of 
spectral functions is the same as for Dirichlet boundary conditions
\eqref{spectra_dir}, which tell us that the free energy has the same
value, $E^c_N=E^c_D$ and $F^{l\not =0}_{U_D}=F^{l\not =0}_{U_D}$. 

\subsection{Periodic boundary conditions}
Periodicity of the fields and anti periodicity of their normal derivatives at the boundaries  
$\varphi(L/2)=\varphi(-L/2)$,$\dot\varphi(L/2)=-\dot\varphi(-L/2)$ correspond to
periodic boundary conditions  associated to the unitary matrix $U_P=\sigma_1$. 
Notice that by definition periodic boundary conditions do relate the boundary 
values of the fields of one boundary with the values of the fields at the other one.

In this case, the derivative of the logarithm of the quotient of spectral functions is
\begin{equation}
	\frac{d}{dk}\log\left(h^L_{U_P}(ik)/h^\infty_{U_P}(ik)\right)=L\coth (kL/2).
\end{equation}

Thus, the Casimir energy is
\begin{align}\label{Cas_Per}
	E^c_{P}(L,m)=-\frac{A}{2\pi L^2}\left(mL\  \text{Li}_2\left(e^{-mL}\right)+\text{Li}_3\left(e^{-mL}\right)\right)
\end{align}
and in the massless limit becomes
\begin{equation}
	E^c_P(L,0)=-\frac{A\zeta(3)}{2\pi L^2}.
\end{equation}
Notice that, in this case the exponential decay of the Casimir energy  $e^{- mL}$ in the asymptotic limit $mL\to \infty$ is slower than that observed in Dirichlet or Neumann boundary conditions, as 
corresponds to the fact that
$\mathrm{Tr}\, \sigma_1 \,\sigma_1= 2\neq 0$.
The rest of terms of the free energy 
\begin{align}\label{Free_Per}
	F^{l\not =0}_{P}(L,m,\beta)=-\frac{AL}{\beta\pi^2}\sum_{l=1}^\infty\int_{m}^\infty dk\ \frac{\sqrt{k^2-m^2}}{l}\Re \left(K_{1}\left(i\beta l\sqrt{k^2-m^2}\right)\right)\left(1-\coth (kL/2)\right).
\end{align}
do share the same behaviour.

\subsection{Anti-Periodic boundary conditions}
Anti-periodic boundary conditions correspond to the values and normal derivatives of the field at the boundary wires satisfying that $\varphi(L/2)=-\varphi(-L/2)$, $\dot\varphi(L/2)=\dot\varphi(-L/2)$, and the associated unitary matrix is $U_A=-\sigma_1$. 
Again in this case, the  boundary conditions relate the boundary values of the fields of one boundary with the boundary values at the other one. In this case the derivative of the logarithm of the quotient of spectral functions is
\begin{equation}
	\frac{d}{dk}\log\left(h^L_{U_A}(ik)/h^\infty_{U_A}(ik)\right)=L\tanh (kL/2).
\end{equation}

Thus, the Casimir energy is
\begin{align}\label{Cas_APer}
	E^c_{U_A}(L,m)=-\frac{A}{2\pi L^2}\left(mL\  \text{Li}_2\left(-e^{-mL}\right)+\text{Li}_3\left(-e^{-mL}\right)\right)
\end{align}
which in the massless limit agrees with the well known results
\begin{equation}
	E^c_A(L,0)=\frac{3A\zeta(3)}{8\pi L^2}.
\end{equation}\label{Free_APer}

Notice that  in this case the exponential decay of the Casimir energy  $e^{- mL}$ is similar to the cae of periodic boundary conditions, as 
corresponds to the fact that
$-\mathrm{Tr}\, \sigma_1 \,\sigma_1= -2\neq 0$.
The rest of the terms of the free energy 
\begin{align} 
	F^{l\not =0}_{U_A}(L,m,\beta)=-\frac{A L}{\beta\pi^2}\sum_{l=1}^\infty\int_{m}^\infty dk\ \frac{\sqrt{k^2-m^2}}{l}\Re \left(K_{1}\left(i\beta l\sqrt{k^2-m^2}\right)\right)\left(1-\tanh(kL/2)\right)
\end{align}
do have the same exponential decay because $1-\tanh(kL/2)\approx e^{-kL}$.
\subsection{Zaremba boundary conditions}
Zaremba boundary conditions correspond to the case where on wire has Dirichlet boundary conditions whereas the other  Neumann boundary conditions. In our parametrization $U_Z=\pm \sigma_3$, and the derivative of the spectral function is 
\begin{equation}
	\frac{d}{dk}\log\left(h^L_{U_Z}(ik)/h^\infty_{U_Z}(ik)\right)=L\tanh (kL).
\end{equation}

The Casimir energy  is
\begin{align}\label{Cas_Zar}
	E^c_{U_Z}(L,m)=-\frac{A}{16\pi L^2}\left(2mL\  \text{Li}_2\left(-e^{-2mL}\right)+\text{Li}_3\left(-e^{-2mL}\right)\right),
\end{align}
that in the massless limit  reduces to
\begin{equation}
	E^c_Z(L,0)=\frac{3 A\zeta(3)}{64\pi L^2}.
\end{equation}

The temperature dependent part of the free energy is 
\begin{equation}\label{Free_Zar}
	F^{l\not =0}_{Z}(L,m,\beta)=-\frac{A L}{\beta\pi^2}\sum_{l=1}^\infty\int_{m}^\infty dk\ \frac{\sqrt{k^2-m^2}}{l}\Re \left(K_{1}\left(i\beta l\sqrt{k^2-m^2}\right)\right)\left(1-\tanh (kL)\right).
\end{equation}

In both cases the exponential suppression  $e^{- 2mL}$ does coincide with that of Dirichlet or Neumann boundary conditions
and again in this case  the  boundary conditions do not relate the boundary values of the fields of one 
boundary with the values at the other one.

\subsection{Asymptotic behaviour}
\begin{figure}[H]
	\includegraphics[width=1\textwidth]{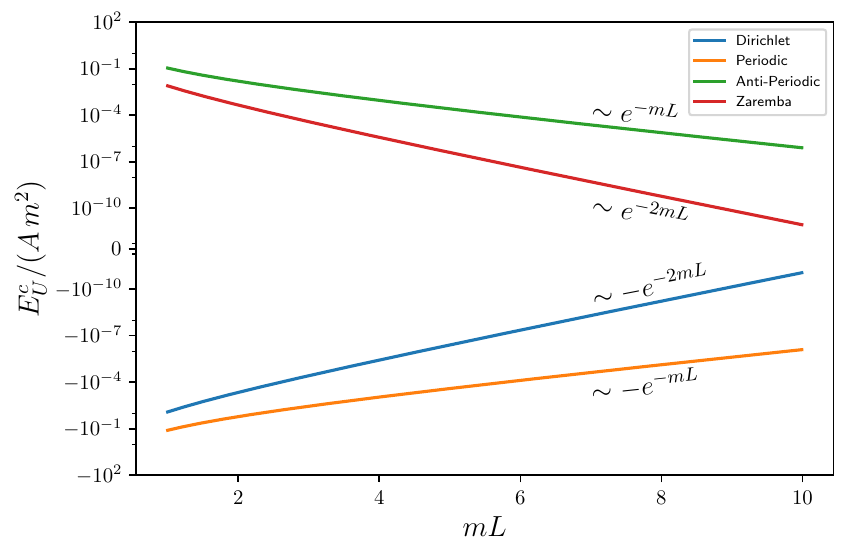}
	\captionof{figure}{Dependence of the Casimir energy 
		in logarithmic scale as a function of the effective distance $mL$ between the two boundary wires for different boundary conditions.}\label{plot_Cas}
\end{figure}

\begin{figure}[H]
	\includegraphics[width=1\textwidth]{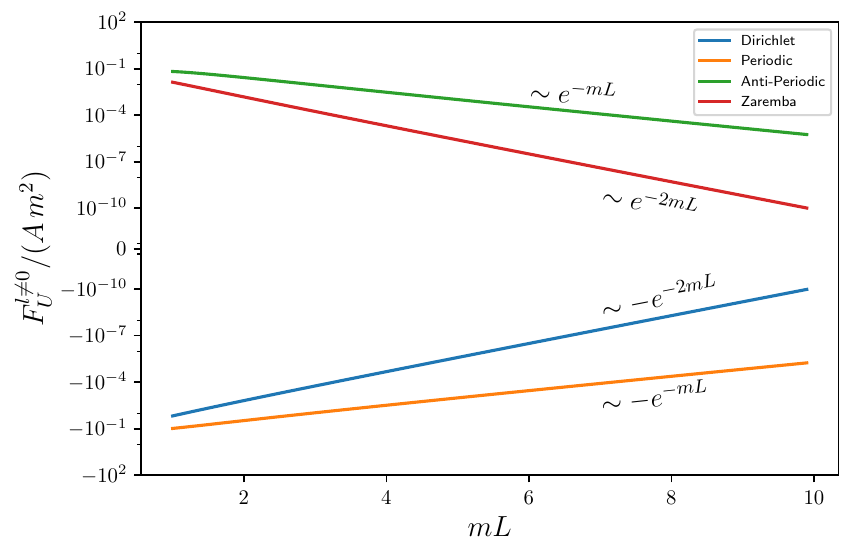}
	\captionof{figure}{Free energy behaviour of the temperature dependent part
		in logarithmic scale as a function of the effective distance $mL$ between the two boundary wires for different boundary conditions, with $m\beta=1$ 
	}
	\label{plot_Free}
\end{figure}

The asymptotic behaviour of the Casimir energy for these boundary conditions follow the rule \eqref{rate} in which  Dirichlet, Neumann \eqref{Cas_Dir} and Zaremba \eqref{Cas_Zar} decays as follows
\begin{equation}
	L E^c_U\sim e^{-2mL}
\end{equation} since for these cases $\text{tr}\left(U\sigma_1\right)=0$, whereas the periodic \eqref{Cas_Per} and anti-periodic \eqref{Cas_APer} behaves like
\begin{equation}
	L E^c_U\sim e^{-mL}
\end{equation} because these boundary conditions satisfy the 
inequality $\text{tr}\left(U\sigma_1\right)\not=0$. We can also appreciate the difference in 
the factor of the exponential decaying behaviour plotting the Casimir energy for 
these boundary conditions (Figure~\ref{plot_Cas}).
By plotting the temperature dependent part of the free energy $F^{l\not=0}_U$ 
(see Figure~\ref{plot_Free}), it can also be seen how these terms exponentially decay to zero as $mL$ grows.
The physical difference between the two families giving rise to different asymptotic behaviours:
the families with faster decay (Dirichlet, Neumann, Zaremba) are conditions imposed on each boundary wire separately, whereas in the second family (periodic, antiperiodic, pseudo periodic) with slower decay rate the boundary conditions involve a relationship between the values of the fields in both wires, establishing a interconnection between them. 
\section{Conclusions}
We have shown the existence of two types of boundary conditions which give rise to different regimes of exponential decay of the Casimir energy at large distances for scalar field theories. The two types are distinguished by the feature that the boundary conditions involve or not inter-connections between the behaviour of the fields at the two boundaries.

The fast exponential decays of the Casimir energy associated to all massive fields makes it negligiable when compared with the contribution of massless fields coming from  electrodynamics. This means that there is no hope of measuring its effects  experimentally. However, from a conceptual viewpoin it can become of crucial  importance to understand the confining infrarred behaviour of non-abelian gauge theories if this regime can be effectively driven by a massive scalar field.

Indeed, analytic arguments \cite{KARABALI1998661,PhysRevD.98.105009} and non-perturbative
numerical simulations \cite{PhysRevLett.121.191601} show that there is a similar exponential decay  in gauge theories with Dirichlet boundary conditions. The verification that such a behaviour is modified for other types of boundary conditions would provide further evidences to the  claim that the low energy behaviour of non-abelian SU(2) gauge theories is governed by an effective scalar field with a fixed non-vanishing mass. The remarkable feature is that the mass of this scalar field is considerably smaller than the lowest   mass of the glueball spectrum \cite{KARABALI1998661,PhysRevD.98.105009}.

In particular the confirmation of the existence of the two regimes for different boundary conditions will be crucial for
the verification of this conjecture.  Numerical simulations are  in progress to clarify this issue.

 \renewcommand{\theequation}{A\arabic{equation}}
\setcounter{equation}{0}  
\begin{appendices}

\section{Alternative calculation of the Free energy}\label{app1}
As an additional check of  previous calculations derived by using the spectral function
of the spatial Laplacian, let us calculate the free energy directly 
in some cases where the spectrum of the spatial Laplacian is explicitly known.
\subsection{Dirichlet boundary conditions}\label{Dir_modes}
In this case the discrete eigenvalues of the spatial Laplacian  are given by $k_j=\pi j/L$ with $j=1,\ldots,\infty$.

Let us consider the {low temperature limit} of the effective action.
The corresponding zeta function \eqref{zeta_2d} is
\begin{equation}
	\zeta(L,m,\beta;s)=\left(\frac{\mu \beta}{2\pi}\right)^{2s}\frac{A\sqrt \pi \Gamma(s-1/2)}{ \beta \Gamma(s)}\sum_{j=1}^\infty \sum_{l=-\infty}^\infty \left(l^2+\left(\frac{j\beta}{2L}\right)^2+\left(\frac{m\beta}{2\pi}\right)^2\right)^{-s+1/2}, 
\end{equation}
that using the Mellin transform \eqref{Mellin_t} and the Poisson formula \eqref{Poisson} on the Matsubara modes we arrive at
\begin{equation} \label{pp}
	\zeta(L,m,\beta;s)=\left(\frac{\mu \beta}{2\pi}\right)^{2s}\frac{A \pi }{ \beta \Gamma(s)}\sum_{j=1}^\infty \sum_{l=-\infty}^\infty\int_{0}^{\infty}dt\ t^{s-2}\ e^{-\left(\left(\frac{j\beta}{2L}\right)^2+\left(\frac{m\beta}{2\pi}\right)^2\right)t-\frac{(\pi l)^2}{t}}.
\end{equation}
After integration, the expression \eqref{pp} reduces to
\begin{align}  \nonumber
	\zeta(L,m,\beta;s)&=\left(\frac{\mu \beta}{2\pi}\right)^{2s}\frac{A\pi}{\beta \Gamma(s)}\left(\Gamma \left(s-1\right)\sum_{j=1}^\infty\left(\left(\frac{j \beta}{2L}\right)^2+\left(\frac{m \beta}{2\pi}\right)^2\right)^{1-s}+\right.\\ \label{LowT_dir_2d}
	&\left. +4\sum_{j,l=1}^\infty\left(\pi l\right)^{s-1}\left(\left(\frac{j \beta}{2L}\right)^2+\left(\frac{m \beta}{2\pi}\right)^2\right)^{1/2-s/2}K_{1-s}\left(\beta l\sqrt{\left(\frac{\pi j}{ L}\right)^2+m^2}\right)\right).
\end{align}

To obtain the contribution of the second term to the effective action  
we have just to derive the gamma function $\Gamma(s)$ on the denominator. We obtain
\begin{equation}
	\left(\zeta^{l\not =0}\right)'(L,m,\beta;0)=\frac{2A}{ L}\sum_{j,l=1}^\infty\frac{1}{l}\sqrt{j^2+\left(\frac{mL}{\pi}\right)^2}K_{1}\left(\frac{\pi\beta l}{L}\sqrt{j^2+\left(\frac{mL}{\pi}\right)^2}\right)
\end{equation}

Now we rewrite  the first term of \eqref{LowT_dir_2d} as
\begin{equation}\label{Dir1}
	\zeta^{l=0}(L,m,\beta;s)=\left(\frac{\mu L}{\pi}\right)^{2s}\frac{A\pi\beta \Gamma \left(s-1\right)}{4L^2 \Gamma(s)}\sum_{j=1}^\infty\left(j^2+\left(\frac{m L}{\pi}\right)^2\right)^{1-s},
\end{equation}
and by applying the Mellin transform \eqref{Mellin_t} and the Poisson formula \eqref{Poisson}, which in this case reads
\begin{equation}\label{Poisson sum_2d}
	\sum_{n=1}^\infty e^{-2\pi \alpha n^2}=\frac{1}{\sqrt{2\alpha }}\sum_{n=1}^{\infty}e^{-\frac{\pi  n^2}{2\alpha}}+\frac{1}{2}\left(\frac{1}{\sqrt{2\alpha }}-1\right),
\end{equation}
for the sum in the modes $j$, we get 
\begin{align} \nonumber
	\zeta^{l=0}(L,m,\beta;s)&=\left(\frac{\mu L}{\pi}\right)^{2s}\frac{A \pi \sqrt \pi \beta}{4L^2 \Gamma(s)}\left(\sum_{j=1}^\infty \int_{0}^{\infty} dt\ t^{s-\frac{5}{2}}e^{-\left(\frac{mL}{\pi}\right)^2t-\frac{(\pi j)^2}{t}}\right.\\ 
	&+\frac{1}{2}\int_{0}^{\infty}dt\ t^{s-\frac{5}{2}}\ e^{-\left(\frac{mL}{\pi}\right)^2t}\left(1-\frac{\sqrt  t}{\sqrt \pi}\right).
\end{align}
After integrating out the $t$ variable we obtain
\begin{align} \nonumber
	\zeta^{l=0}(L,m,\beta;s)&=\left(\frac{\mu L}{\pi}\right)^{2s}\frac{A \pi \sqrt \pi \beta}{8L^2 \Gamma(s)}\left(\Gamma \left(s-\frac{3}{2}\right)\left(\frac{mL}{\pi}\right)^{3-2s}-\frac{\Gamma(s-1)}{\sqrt \pi }\left(\frac{mL}{\pi}\right)^{2-2s}\right.\\ 
	&+4\sum_{j=1}^\infty \left(\pi j\right)^{s-3/2}\left(\frac{mL}{\pi}\right)^{3/2-s} K_{3/2-s}\left(2jmL\right).
\end{align}

Upon derivation the only non-vanishing contribution of  this term comes from the derivative of $\Gamma(s)$ in the first and third term, whereas we have to use the asymptotic expansion \eqref{s_expansion_G} for the second. The result is
\begin{align}\nonumber
	\left(\zeta^{l=0}\right)'(L,m,\beta;0)=&\frac{AL\beta m^3}{6\pi}+\frac{Am^2\beta}{4\pi}\left(\log (\mu/m)+\frac{1}{2}\right)\\ \label{Dir_f}
	&+\frac{A\beta}{8L^2\pi}\left(2mL\ \text{Li}_2\left(e^{-2mL}\right)+\text{Li}_3\left(e^{-2mL}\right)\right).
\end{align}

From the renormalized effective action \eqref{zeta_combination} 
we can compute the Casimir energy 
\begin{equation}
	E^c_D(L,m)=-\frac{A}{16L^2\pi}\left(2mL\ \text{Li}_2\left(e^{-2mL}\right)+\text{Li}_3\left(e^{-2mL}\right)\right)
\end{equation}
which is the same that we obtained in \eqref{Cas_Dir} with the spectral function.
The temperature dependent component of the free energy
\begin{align}
	F^{l\not =0}_D(L,m,\beta)=&-\frac{A}{ \beta L}\sum_{j,l=1}^\infty\frac{1}{l}\sqrt{j^2+\left(\frac{mL}{\pi}\right)^2}K_{1}\left(\frac{\pi\beta l}{L}\sqrt{j^2+\left(\frac{mL}{\pi}\right)^2}\right)\\
	&-\frac{1} \nonumber {2\beta}\lim_{L_0\rightarrow\infty}\left(\left(\zeta^{l\not =0}\right)'(2L_0+L,m,\beta;0)-2\left(\zeta^{l\not =0}\right)'(L_0+L,m,\beta;0)\right).
\end{align}
that can be shown to be  equivalent to \eqref{Free_Dir}.

\subsection{Periodic boundary conditions}\label{Per_modes}

The discrete eigenvalues of the spatial Laplacian in this case  are $k_j=2\pi j/L$ with $j\in \mathbb{Z}$. We can derive the effective action directly from the spectrum
by rewriting \eqref{ini_1}
\begin{equation}
	\zeta(L,m,\beta;s)=\left(\frac{\mu \beta}{2\pi}\right)^{2s}\frac{A\sqrt \pi \Gamma(s-1/2)}{\Gamma(s)}\sum_{l,j=-\infty}^\infty \left(l^2+\left(\frac{j\beta}{L}\right)^2+\left(\frac{m\beta}{2\pi}\right)\right)^{-s+1/2},
\end{equation}
and by using the Mellin transform and the Poisson formula on the Matsubara modes, we have
\begin{equation}
	\zeta(L,m,\beta;s)=\left(\frac{\mu \beta}{2\pi}\right)^{2s}\frac{A \pi }{\Gamma(s)}\sum_{l,j=-\infty}^\infty \int_{0}^\infty dt\ t^{s-2}e^{-\left(\left(\frac{j\beta}{L}\right)^2+\left(\frac{m\beta}{2\pi}\right)\right)t-\frac{(\pi l)^2}{t}}	,
\end{equation}
which, after integration, becomes
\begin{align}\label{lowT_2d_ini}
	\zeta(L,m,\beta;s)&=\left(\frac{\mu \beta}{2\pi}\right)^{2s}\frac{A\pi}{\beta \Gamma(s)}\sum_{j=-\infty}^\infty 
	\left(\Gamma(s-1)
	\left(\left(\frac{j\beta}{L}\right)^2+\left(\frac{m\beta}{2\pi}\right)^2\right)^{1-s}\right.\\ \nonumber
	&\left.+4\sum_{l=-\infty}^\infty \left(\pi l\right)^{s-1}\left(\left(\frac{j\beta}{L}\right)^2+\left(\frac{m\beta}{2\pi}\right)^2\right)^{1/2-s/2} K_{1-s}\left(\beta l \sqrt{\left(\frac{2\pi j}{ L}\right)^2+m^2}\right)\right)\textcolor{red}{.}
\end{align}

The derivative of the second term gives
\begin{equation}
	\left(\zeta^{l\not =0}\right)'(L,m,\beta;0)=\frac{4A}{L}\sum_{j=-\infty}^\infty\sum_{l=1}^\infty \frac{1}{l}\sqrt{j^2+\left(\frac{mL}{2\pi}\right)^2} K_{1}\left(\frac{2\pi\beta l}{L} \sqrt{j^2+\left(\frac{mL}{2\pi}\right)^2}\right).
\end{equation}

We rewrite \eqref{lowT_2d_ini} as
\begin{equation}
	\zeta^{l=0}(L,m,\beta;s)=\left(\frac{\mu L}{2\pi}\right)^{2s}\frac{A\beta\Gamma(s-1)\pi}{L^2 \Gamma(s)}\sum_{j=-\infty}^\infty \left(j+\left(\frac{mL}{2\pi }\right)^2\right)^{1-s}
\end{equation}
and follow the same strategy as for  Dirichlet boundary conditions from equation \eqref{Dir1} to \eqref{Dir_f} with these particular spatial modes. Thus, we arrive at 
\begin{align}
	\left(\zeta^{l =0}\right)'(L,m,\beta;0)=\frac{AL\beta m^3}{6\pi}+\frac{A\beta}{L^2\pi}\left(mL\  \text{Li}_2\left(e^{-mL}\right)+\text{Li}_3\left(e^{-mL}\right)\right)\textcolor{red}{}
\end{align}
and the Casimir energy is
\begin{equation}
	E^c_P(L,m)=-\frac{A}{2\pi L^2}\left(mL\  \text{Li}_2\left(e^{-mL}\right)+\text{Li}_3\left(e^{-mL}\right)\right)
\end{equation} which does coincide with result obtained by the general spectral function method \eqref{Cas_Per}. Meanwhile, the temperature dependent component of the free energy is
\begin{align}\nonumber
	F_P^{l\not =0}(L,m,\beta)=&-\frac{2A}{\beta L}\sum_{j=-\infty}^\infty\sum_{l=1}^\infty \frac{1}{l}\sqrt{j^2+\left(\frac{mL}{2\pi}\right)^2} K_{1}\left(\frac{2\pi\beta l}{L} \sqrt{j^2+\left(\frac{mL}{2\pi}\right)^2}\right)\\
	&-\frac{1}{2\beta}\lim_{L_0\rightarrow\infty}\left(\left(\zeta^{l\not =0}\right)'(2L_0+L,m,\beta;0)-2\left(\zeta^{l\not =0}\right)'(L_0+L,m,\beta;0)\right)
\end{align}which does also agree with \eqref{Free_Per}.
\end{appendices}

\newpage
\section*{Acknowledgements}
We thank  J.M.  Mu\~noz-Casta\~neda and V.P. Nair  for discussions. C.I. is grateful to the University of Zaragoza for the hospitality.
We are partially supported by Spanish Grants No. PGC2022-126078NB-C21 funded by MCIN/AEI/10.13039/ 501100011033, Erasmus+ Programme, ERDF A way of making EuropeGrant; the Quantum Spain project of the QUANTUM ENIA of the Ministry of Economic Affairs and Digital Transformation, the Diputación General de Aragón-Fondo Social Europeo (DGA-FSE) Grant No. 2020-E21-17R of the Aragon Government, and the European Union, NextGenerationEU Recovery and Resilience Program on 'Astrof\ii sica y F\ii sica de Altas Energ\ii as , CEFCA-CAPA-ITAINNOVA.


\renewcommand{\refname}{References}
\bibliographystyle{unsrt}
\bibliography{refe}

\begin{thebibliography}{10}

\bibitem{Casimir:1948dh}
H.~B.~G. Casimir.
\newblock {On the attraction between two perfectly conducting plates}.
\newblock {\em Indagationes Mathematicae}, 10(4):261--263, 1948.

\bibitem{sparnaay1957attractive}
MJ~Sparnaay.
\newblock Attractive forces between flat plates.
\newblock {\em Nature}, 180(4581):334--335, 1957.

\bibitem{sparnaay1958measurements}
M~J Sparnaay.
\newblock Measurements of attractive forces between flat plates.
\newblock {\em Physica}, 24(6-10):751--764, 1958.

\bibitem{lamoreaux1997demonstration}
Steven~K Lamoreaux.
\newblock Demonstration of the casimir force in the 0.6 to 6 $\mu$ m range.
\newblock {\em Physical Review Letters}, 78(1):5, 1997.

\bibitem{PhysRevLett.81.4549}
U.~Mohideen and Anushree Roy.
\newblock Precision measurement of the casimir force from 0.1 to
  $0.9\mathit{\ensuremath{\mu}}m$.
\newblock {\em Phys. Rev. Lett.}, 81:4549--4552, Nov 1998.

\bibitem{PhysRevLett.87.211801}
H.~B. Chan, V.~A. Aksyuk, D.~J. Kleiman, R. N.~Bishop, and Federico Capasso.
\newblock Nonlinear micromechanical casimir oscillator.
\newblock {\em Phys. Rev. Lett.}, 87:211801, Oct 2001.

\bibitem{Dalvit}
D~Dalvit, P~Milonni, D~Roberts, and F.~da~Rosa.
\newblock Casimir physics.
\newblock {\em Lecture Notes in Physics}, 834, June 2011.

\bibitem{Boundary_general_2013}
M.~Asorey and Jose Muñoz-Castañeda.
\newblock Attractive and repulsive casimir vacuum energy with general boundary
  conditions.
\newblock {\em Nuclear Physics B}, 874, 06 2013.

\bibitem{munoz2020thermal}
JM~Mu{\~n}oz-Casta{\~n}eda, L~Santamar{\'\i}a-Sanz, M~Donaire, and
  M~Tello-Fraile.
\newblock Thermal casimir effect with general boundary conditions.
\newblock {\em The European Physical Journal C}, 80:1--16, 2020.

\bibitem{SYMANZIK19811}
K.~Symanzik.
\newblock Schrodinger representation and casimir effect in renormalizable
  quantum field theory.
\newblock {\em Nuclear Physics B}, 190(1):1--44, 1981.

\bibitem{KARABALI1998661}
C.~Karabali, D.~Kim and V.P. Nair.
\newblock Planar yang-mills theory: Hamiltonian, regulators and mass gap.
\newblock {\em Nuclear Physics B}, 524(3):661--694, 1998.

\bibitem{PhysRevD.98.105009}
D.~Karabali and V.~P. Nair.
\newblock Casimir effect in ($2+1$)-dimensional yang-mills theory as a probe of
  the magnetic mass.
\newblock {\em Physical Review D}, 98:105009, Nov 2018.

\bibitem{karabali1996origin}
Dimitra Karabali and VP~Nair.
\newblock On the origin of the mass gap for non-abelian gauge theories in (2+
  1) dimensions.
\newblock {\em Physics Letters B}, 1(379):141--147, 1996.

\bibitem{PhysRevLett.121.191601}
M.~N. Chernodub, V.~A. Goy, A.~V. Molochkov, and Ha~Huu Nguyen.
\newblock Casimir effect in yang-mills theory in $d=2+1$.
\newblock {\em Physical Review Letters}, 121:191601, Nov 2018.

\bibitem{asorey2023energy}
M.~Asorey and F.~Ezquerro.
\newblock Energy preserving boundary conditions in field theory.
\newblock {\em Physical Review D}, 4:045008, 2023.

\bibitem{asorey_temp1}
M.~Asorey, C.~Beneventano, D.~D'Ascanio, and Eve Santangelo.
\newblock Thermodynamics of conformal fields in topologically non-trivial
  space-time backgrounds.
\newblock {\em Journal of High Energy Physics}, 2013, 12 2012.

\bibitem{asorey2015topological}
M~Asorey, C~G Beneventano, I~Cavero-Pel{\'a}ez, Daniela D’Ascanio, and E~M
  Santangelo.
\newblock Topological entropy and renormalization group flow in 3-dimensional
  spherical spaces.
\newblock {\em Journal of High Energy Physics}, 2015(1):1--35, 2015.

\bibitem{zou2013casimir}
J.~{\sl et al} Zou.
\newblock Casimir forces on a silicon micromechanical chip.
\newblock {\em Nature communications}, 4(1):1845, 2013.

\bibitem{terccas2017quantum}
H.~{\sl et al} Ter{\c{c}}as.
\newblock Quantum thermal machines driven by vacuum forces.
\newblock {\em Physical Review E}, 95(2):022135, 2017.

\bibitem{geyer2005thermal}
B.~Geyer, G.L. Klimchitskaya, and V.M. Mostepanenko.
\newblock Thermal corrections in the casimir interaction between a metal and
  dielectric.
\newblock {\em Physical Review A}, 72(2):022111, 2005.

\bibitem{milton2017casimir}
P.~Milton, K. A~Kalauni, P.~Parashar, and Y.~Li.
\newblock Casimir self-entropy of a spherical electromagnetic $\delta$-function
  shell.
\newblock {\em Physical Review D}, 96(8):085007, 2017.

\bibitem{bordag2018free}
M.~Bordag.
\newblock Free energy and entropy for thin sheets.
\newblock {\em Physical Review D}, 98(8):085010, 2018.

\bibitem{asorey2005global}
M~Asorey, A.~Ibort, and G.~Marmo.
\newblock Global theory of quantum boundary conditions and topology change.
\newblock {\em International Journal of Modern Physics A}, 20(05):1001--1025,
  2005.

\bibitem{PhysRevD.13.3224}
J.~S. Dowker and R.~Critchley.
\newblock Effective lagrangian and energy-momentum tensor in de sitter space.
\newblock {\em Physical Review D}, 13:3224--3232, Jun 1976.

\bibitem{Blau:1988kv}
S.~Blau, M.~Visser, and A.~Wipf.
\newblock {Zeta Functions and the Casimir Energy}.
\newblock {\em Nuclear Physics B}, 310:163, 1988.

\bibitem{PhysRevD.56.7797}
Valter Moretti.
\newblock Direct $\ensuremath{\zeta}$-function approach and renormalization of
  one-loop stress tensors in curved spacetimes.
\newblock {\em Phys. Rev. D}, 56:7797--7819, Dec 1997.

\bibitem{10.1063/1.532929}
Valter Moretti.
\newblock {One-loop stress-tensor renormalization in curved background: The
  relation between $\zeta$-function and point-splitting approaches, and an
  improved point-splitting procedure}.
\newblock {\em Journal of Mathematical Physics}, 40(8):3843--3875, 08 1999.

\end{thebibliography}

\end{document}